\documentclass[%
 reprint,
superscriptaddress,
amsmath,amssymb,
aps,
pra,
]{revtex4-1}
\UseRawInputEncoding
\usepackage{caption}
\usepackage{subcaption}
\usepackage{graphicx}
\usepackage{xcolor}
\usepackage{graphicx}
\usepackage{dcolumn}
\usepackage{bm}
\usepackage{here}
\usepackage{comment}
\usepackage[normalem]{ulem}
\usepackage{color}

\DeclareRobustCommand{\erase}{\bgroup\markoverwith{\textcolor{red}{\rule[.5ex]{2pt}{0.4pt}}}\ULon}

\begin{document}

\title{Experimental Validation of String Oscillation in Subharmonic Generation}
\author{Shotaro Kawano}
\affiliation{Department of Physics, Graduate School of Science, The University of Tokyo, 7-3-1 Hongo, Bunkyo-ku, Tokyo 113–0033, Japan}
\author{Kenji Kobayashi}
\affiliation{Graduate School of Pure and Applied Sciences, University of Tsukuba, 1-1-1 Tennodai, Tsukuba-shi, Ibaraki 305-8571, Japan}
\author{Takuya Suzuki}
\affiliation{Graduate School of Pure and Applied Sciences, University of Tsukuba, 1-1-1 Tennodai, Tsukuba-shi, Ibaraki 305-8571, Japan}
\author{Naoki Ichiji}
\email{ichiji@iis.u-tokyo.ac.jp}
\affiliation{Institute of Industrial Science, The University of Tokyo, 4-6-1 Komaba, Meguro-Ku, Tokyo 153-8505, Japan}

\begin{abstract}
The lowest notes produced by string instruments are typically limited by the fundamental vibration of the strings. 
However, precise control of bow pressure can lead to the production of even lower notes. 
Despite significant interest in this counterintuitive technique and various proposed explanations, no conclusive evidence has been provided, making detailed discussions of the underlying mechanism challenging.
In this study, we employ high-speed imaging to visualize the spatial vibration modes of stringed instruments, confirming Helmholtz motion and its modifications under subharmonic conditions. 
Finite element simulations further demonstrated that increased bow pressure amplifies frictional forces, suppressing standard vibrations and allowing subharmonic frequencies to emerge. 
Our results provide the clear experimental validation of the mechanism underlying subharmonic sound production, providing an avenue for further exploration of vibrational and oscillatory phenomena.
\end{abstract}

\maketitle

\begin{figure}[b]
  \begin{center}
  \includegraphics[width=8.6cm]{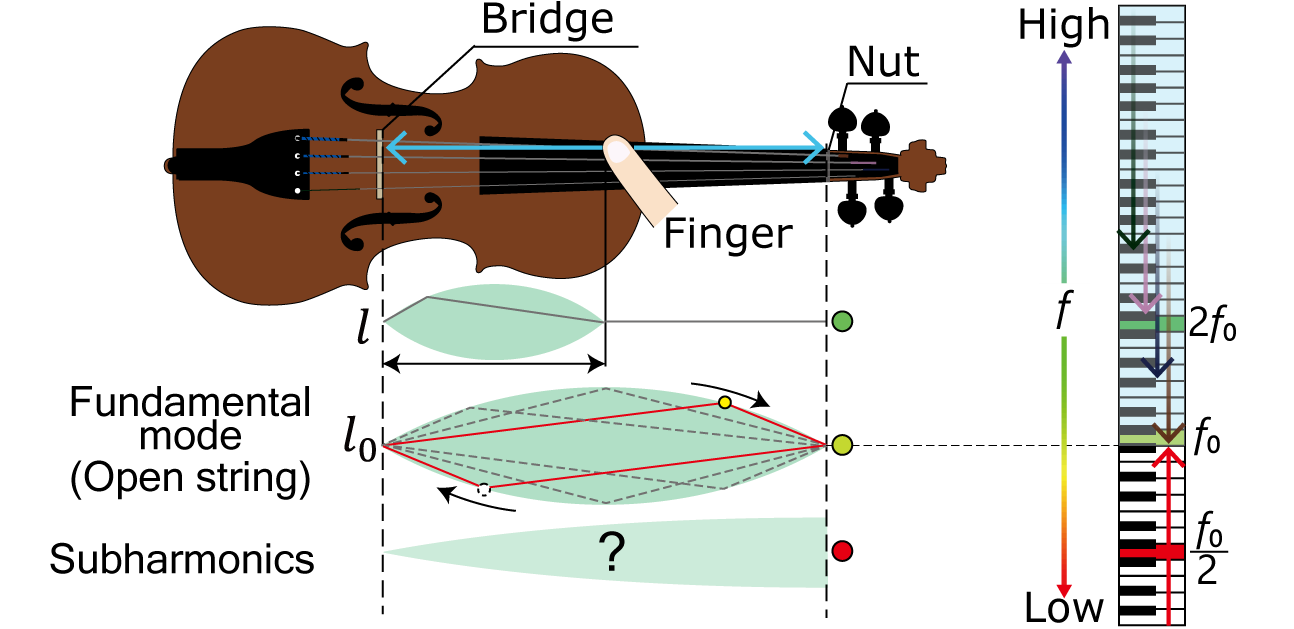}
  \end{center}
  \caption{Schematic diagram of the tonal mechanism of a violin. The keyboard on the right side shows the schematics of the frequency range. The translucent light blue area and the red marker indicate the ranges that can be played with normal techniques and subharmonics, respectively.}
  \label{Fig:image}
\end{figure}

One of the defining features of string instruments in the violin family is their capacity to continuously modulate pitch across a wide range.
The string vibrations in a violin exhibit a complex motion known as Helmholtz motion~\cite{Helmholtz1954,Wood04ACTA}, where a folded point travels along the string envelope, as schematically illustrated in Fig.~\ref{Fig:image}(a). The perceived pitch is determined by the fundamental frequency $f\!=\!\tfrac{1}{2l}\sqrt{S/\rho}$, where $l$, $S$, and $\rho$ denote the vibrating length, tension, and line density of the string, respectively. 
Since the line density of each string is fixed and adjusting tension during performance is impractical, pitch modulation is generally achieved by controlling the vibrating length ($l$) by pressing on the string with the fingers (Fig.~\ref{Fig:image}(b)).

As $f$ is inversely proportional to $l$, shorter vibrating lengths generate higher frequencies, theoretically allowing for an unlimited upper range. 
In contrast, the lower frequency limit is determined by the maximum vibrating length $l_{0}$, corresponding to the distance between the bridge and nut, with the lowest frequency given by $f_{0}\!=\!\tfrac{1}{2l_{0}}\sqrt{S/\rho}$.

However, violinist M. Kimura demonstrated an extended technique that enables the production of sounds approximately one octave below $f_{0}$~\cite{Guettler1994CASJ,Hanson95ASA}.
This technique, known as subharmonics or Anomalous Low Frequency (ALF), is achieved by applying consistent and intense bow pressure to the string~\cite{Kimura99NMR}.
The generation of frequencies below the fundamental vibration—a seemingly counterintuitive and thus intriguing phenomenon—has attracted the attention of both musicians and scientists. This phenomenon has been the subject of numerous studies in physics, with several reports providing qualitative explanations of its underlying principles~\cite{RossingBook,Schoonderwaldt09ACTA}.
However, experimental validation has remained limited~\cite{Hanson95ASA,Yoshikawa1997}, and the absence of detailed verification of two-dimensional string vibration modes during subharmonic production has hindered further understanding of the low-frequency generation process. Consequently, subharmonics continue to be perceived as an unexplained phenomenon despite substantial prior research.

In this study, we directly visualize the spatial vibration modes of strings during subharmonic production by capturing high-speed images of the strings during actual performances. By comparing the two-dimensional behavior of the strings under both normal and subharmonic techniques, we unambiguously demonstrate the mechanism by which vibrations below the fundamental frequency are generated. This visualization of dynamic vibration modes enables a more detailed examination of the underlying mechanisms. Furthermore, numerical simulations based on the experimental conditions confirms that precise control of bow pressure and bowing speed is crucial for achieving subharmonics.

\begin{figure}[t!]
\centering
  \includegraphics[width=8.6cm]{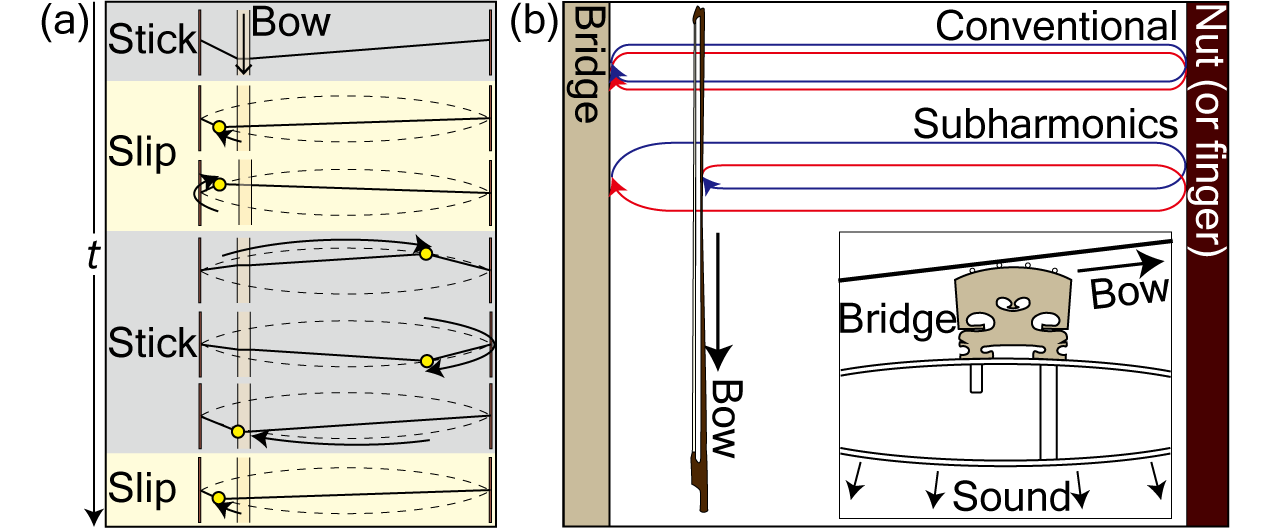}
\caption{Schematic of the (a) Helmholtz motion and (b) the proposed explanation for the appearance of subharmonics~\cite{Guettler02}. The red and blue arrows indicate the first and second cycles of the circumferential motion of bending angles, respectively. The inset shows a schematic cross-section at the position of the bridge, illustrating the transmission of string vibrations.}
\label{Fig:spatialmode}
\end{figure}

First, we briefly describe the sounding mechanism of bowed instruments.
Sound production in a bowed instrument is primarily driven by periodic stick-slip motion at the bow-string interface~\cite{Wood04ACTA, Akar20Vib, Raman1918}.
Static friction initially causes the string to stick to the bow and deform in the direction of bow movement. When the tension at the contact point exceeds the static friction force, the string slips, repeating in cycles that generate a vibratory motion of the string.
This stick-slip mechanism creates a bending angle (Helmholtz angle) that propagates along the string, reflects at the fixed ends, and completes a lens-shaped circulating path, as illustrated in Fig.~\ref{Fig:image}(a). The returning bending angle triggers the next slip event, maintaining a stable vibration mode at a constant frequency, even with variations in bow speed and pressure.
Vibrations then travel through the bridge (Fig.~2(b), inset), exciting the instrument body and emitting sound. Since the sound is primarily determined by the vibrations reaching the bridge, any alteration in how these vibrations are transmitted can significantly affect the perceived pitch and tone.

Under normal playing, the Helmholtz angle reflects once at each end per cycle, completing a circulating vibration pattern. In contrast, in subharmonic playing~\cite{FletcherBook,Hanson95ASA,Guettler02,Schoonderwaldt09ACTA}, the high bow pressure required for subharmonics increases friction at the contact point, making slip motion more difficult compared to normal playing. As a result, the contact point behaves as an additional reflective boundary. When this additional reflection occurs once every two cycles, the rate at which the bridge receives the Helmholtz corner is halved, effectively reducing the frequency transmitted from the string vibration to the instrument to half.

\begin{figure*}[t!]
\begin{center}
\includegraphics[width=17cm]{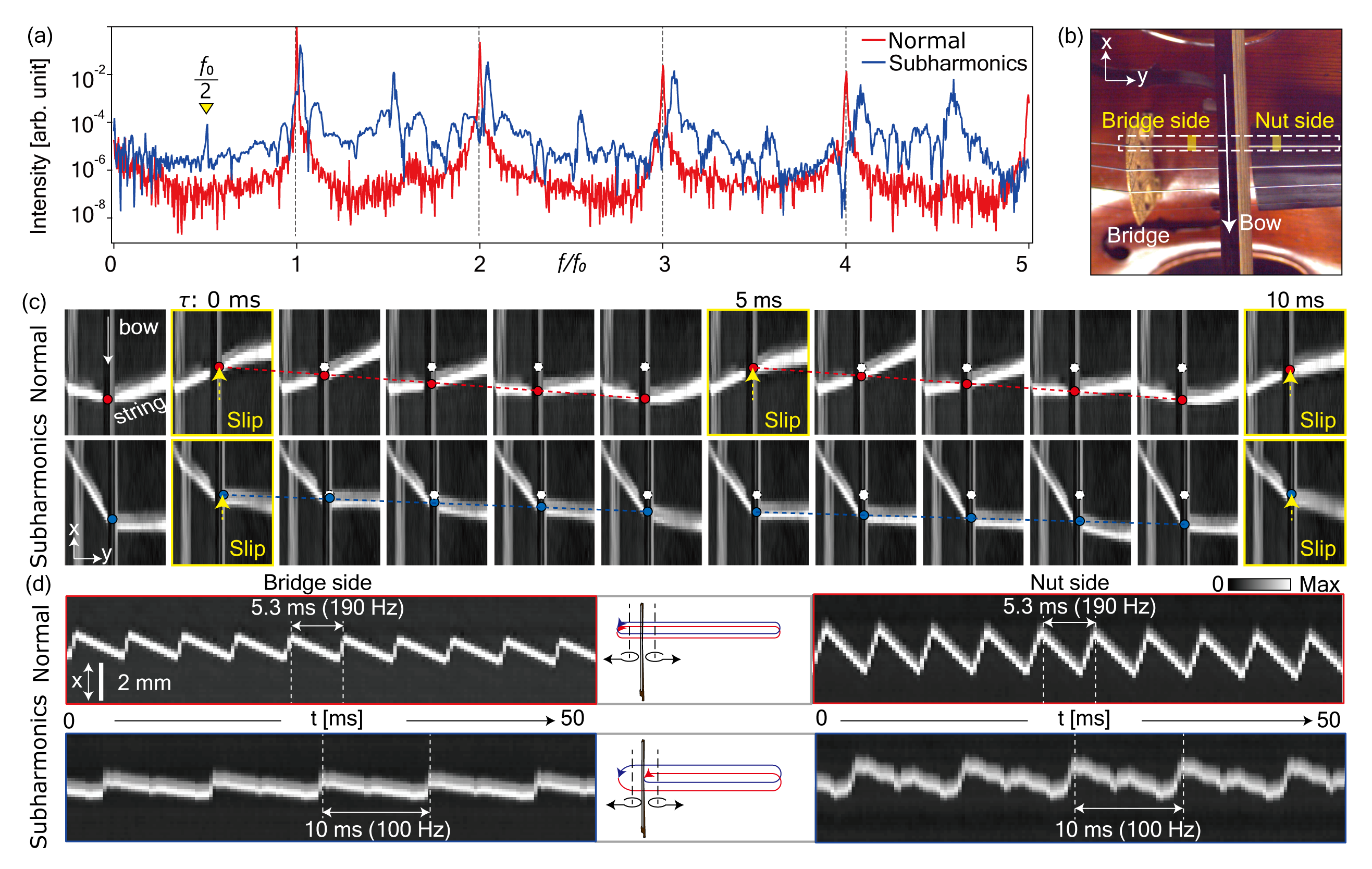}
\caption{(a) FFT spectrum comparing the normal playing technique (red) and subharmonic technique (blue).
(b) A frame of high-speed movie showing the setup for analyzing string vibrations. (c) Representative frames of high-speed movies plotted at 1 ms intervals. The contact points between the bow and the string are indicated by colored markers, and the motion of the contact points during the stick motion is indicated by colored dashed lines. Frames in which slip motion occurs are highlighted with yellow frames. (d) Two-dimensional plot of the temporal variation of the string. The regions indicated by the translucent yellow bands in (b) are integrated in the $y$-direction and plotted with time on the horizontal axis. }
\label{Fig:exp}
\end{center}
\end{figure*}

Historically, the study of string instruments relied on electromagnetic induction recording, which effectively captured velocity but posed challenges in directly observing the two-dimensional deformation of the string during vibration~\cite{Schoonderwaldt2008, Schoonderwaldt09ACTA, Cremer1984}. With advancements in high-speed camera technology, the ability to visualize two-dimensional motions in the audible range has significantly improved, enabling a more detailed analysis of sound generation mechanisms across various research fields~\cite{Phillips2018, Hack2012, Cook24PNAS}.
In this study, we experimentally investigate the proposed mechanism by comparing string vibrations during performance of normal and subharmonic techniques using a high-speed camera under performance conditions on an actual instrument.

While the range of pitches generated through subharmonics can be tuned by varying the bow position and applied pressure~\cite{Kimura99NMR}, this study focuses on the simplest mode, producing a sound approximately one octave lower than the fundamental frequency to demonstrate the underlying principle.
Fig.~\ref{Fig:exp}(a) shows the FFT spectrum obtained by positioning the bow approximately 5 cm from the bridge, applying strong vertical pressure, and moving the bow slowly. The string used is a Pirastro Violino (G string).
The abscissa represents frequencies normalized to the fundamental frequency of the string. For comparison, the spectrum for the normal technique (red) is also plotted, showing peaks exclusively at integer multiples of $f_{0}$, which represent the harmonic series based on the fundamental frequency.
In contrast, the spectrum for the subharmonic technique (blue) shows distinct peaks at $n \cdot (f_{0}/2)$, with the presence of $f_{0}/2$ providing clear evidence of modes below the fundamental frequency.

The string motions during each technique were captured using a high-speed camera (Photoron, FASTCAM-APX RS) with a shutter speed of 1/3000 s and a frame rate of 3000 fps. 
Fig.~\ref{Fig:exp}(b) shows one of the acquired frames, while Fig.~\ref{Fig:exp}(c) presents a typical vibration cycle, with the aspect ratio adjusted to enhance the visibility of the minute spatial amplitudes of the string vibrations. The plotted regions were cropped based on the bow-string contact point, with the white dashed area in (b) corresponding to the region shown for the normal technique. For the subharmonic technique, the cropping was similarly centered around the contact point to capture its distinct motion. The moment a specific slip event occurred was designated as $\tau\!=\!0$, with subsequent frames shown at 1 ms intervals up to $\tau\!=\!11$ ms.

The red and blue markers indicate the contact points between the string and the bow, while the white markers show the positions of the contact points immediately after each slip event.
The gradual downward shift of the contact point followed by its momentary restoration clearly confirms the stick-slip phenomenon.
The slip event, highlighted by the yellow frames, occurs every 5 ms in the normal technique (top panel) and at half that frequency in the subharmonic technique (bottom panel).

Upon closer inspection, slight perturbations of the contact point can be observed in the subharmonic technique at frames corresponding to $\tau\!=\!5$–$6$ ms, when slips occur in the normal technique due to the returning Helmholtz angle.
In the subharmonic technique, the strong pressure applied to the string evident from the angle between the string and bridge shown in Fig.~\ref{Fig:exp}(c) generates a greater static frictional force, preventing the Helmholtz angle collision from triggering the next slip event.
Another contributing factor is the slow bow speed, which reduces the incremental change in tension per cycle, resulting in insufficient tension buildup at the contact point.
As a result, no slip phenomenon occurs upon the first arrival of the Helmholtz angle, and the string remains in stick motion.
When the Helmholtz angle reaches the string for the second time and the deformation of the string has sufficiently accumulated, the slip phenomenon occurs, resulting in the frequency perceived at the bridge being reduced to half.

To facilitate a clearer comparison of the temporal behavior across the bridge, we plotted the time variation of the $x$ displacement of the string as two-dimensional color maps in Figs.~\ref{Fig:exp}(d). The regions located at a distance of ±2 cm from the contact point toward both the nut and bridge sides, corresponding to the areas highlighted by the translucent yellow squares in Fig.~\ref{Fig:exp}(c), were integrated along the $y$-direction and plotted with elapsed time on the horizontal axis.
Since the peak intensity of the image correlates with the string position, the sawtooth waveforms in Figs.~\ref{Fig:exp}(d) directly represent the time variations in the string oscillation along the $x$-direction.
The asymmetrical sawtooth waveform, characterized by a gradual downward shift followed by a rapid return, clearly illustrates the stick-slip phenomenon, consistent with various previous studies~\cite{Cremer1984, RossingBook}.

In the normal playing technique (top panels), the waveforms on both the bridge side and the nut side exhibit similar shapes, differing primarily in amplitude.
In contrast, the string vibrations in the subharmonic technique (bottom panels) show distinctly different behaviors on the two sides of the bow. 
The waveform on the bridge side resembles that of the normal technique but is stretched horizontally by a factor of two, indicating that the vibration reaches the bridge side only once for every two cycles in comparison to the normal technique.
On the nut side, however, the waveform exhibits undulating patterns with central distortion, along with sawtooth waves occurring at the same intervals as those on the bridge side, indicating the presence of both the reference frequency and its half-frequency component.
Although slight distortion is visible at the corresponding position on the bridge side, the difference is more pronounced on the nut side.
These observations suggest that the fundamental frequency component does not propagate to the bridge side but remains confined to the nut side due to the reflection of the first-round Helmholtz angle at the contact point.

Based on these experimental results, we conclude that the observed waveform features provide experimental evidence of the Helmholtz motion reflecting at the bow-string contact point every two cycles during subharmonics, consistent with previous theoretical explanations of the subharmonics technique~\cite{Guettler02}. Furthermore, the two-dimensional spatial distribution of string vibrations and the temporal variation of displacements captured in our experiments suggest that bow pressure and bow speed are the dominant factors.
However, the information available from the captured images remains limited and insufficient to fully identify the dominant mechanisms of the stick-slip phenomenon, which is a highly complex process. 
To gain deeper insight into the underlying physical factors, we further investigate the key mechanisms governing subharmonics through numerical finite element (FE) simulations \cite{Akar2020-tg, Bellante2015-si}, focusing on frictional forces as the external driving force.

\begin{figure}[t!]
\centering
  \includegraphics[width=8.6cm]{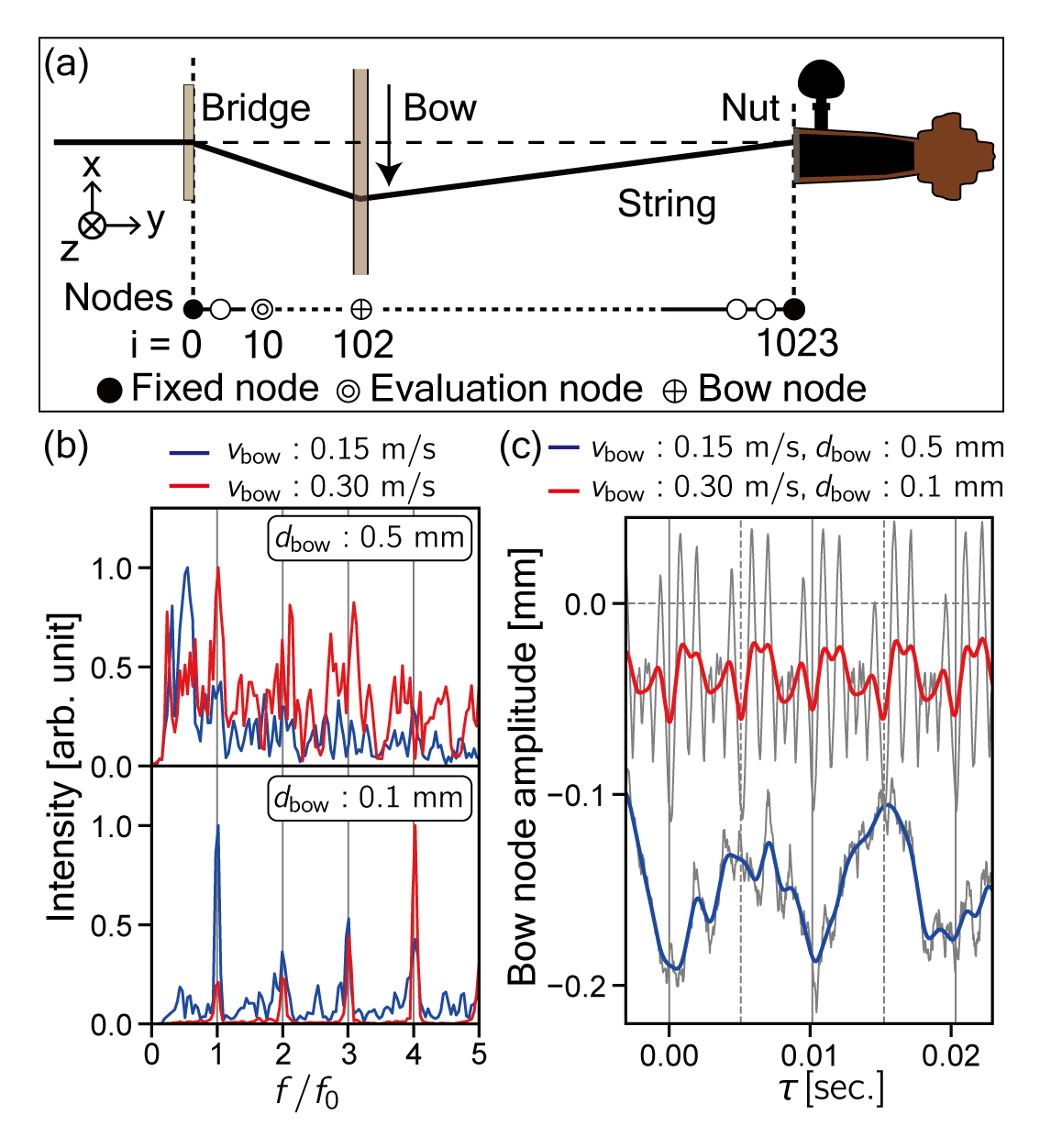}
\caption{(a) Schematic of the components of a string instrument and the discretized string model. The bow and string interact only at a single point on the bow node. The sound waves transmitted from the bridge to the body are calculated at the evaluation node closest to the bridge. (b) Spectra at the evaluation node corresponding to different pressing depths $(d_{\mathrm{bow}})$ and bow speeds $(v_{\mathrm{bow}})$. (c) Time variations in the amplitude of the bow node where the pressing depth is deep and the bow speed is slow (blue) and where the pressing depth is shallow, and the bow speed is fast (red). The colored waves are the low-pass filtered result of the original waveform (black line) using a Hanning window.}
\label{Fig:analysis}
\end{figure}

The simple model from Reference~\cite{Bellante2015-si} was employed for the numerical simulation.
Fig.~\ref{Fig:analysis}(a) describes the components of a string instrument and the discretized string consisting of a finite number of nodes. 
The coordinate set is a right-handed system with the $y$-axis in the direction of the string. 
The black nodes on the bridge and nut are fixed. 
The cross-marked node, called the ``bow node'', represents the contact position with the bow.
Each unfixed node interacts with both neighbors according to the following equation of motion: 
\begin{align}
\label{eq:eom_no_bow2}
dm \frac{\partial^2 \bm{u}(x_0,t)}{\partial t^2}=AEdx_0\frac{\partial}{\partial x_0}\left( \varepsilon(x_0,t)\bm{t}(x_0,t) \right), 
\end{align}
where $dm$ is the line element mass, $A$ is the cross-section area of the string, $E$ is the Young's modulus, $\varepsilon(x_0,t)$ is the true strain, and $\bm{t}$ is a tangent vector along the edge. 
The acceleration of the bow node is modified by the relative motion between the string and the bow. 
In stick motion, there is no acceleration because the bow node follows the bow at a constant velocity.

During slip motion, the string slips against the bow, and the acceleration is modified by the dynamic frictional force according to its relative velocity. 
Thus, the acceleration of the bow node in the $y$-direction is modified as, 
\begin{align}
\label{eq:slip_acc}
\alpha_y^{\rm bow}=\alpha^f_y+\mu' \left| \alpha^f_z \right|,
\end{align}
whrere $\alpha^f$ is the acceleration field in the case of the free motion of the string, and $\mu'$ is the dynamic friction coefficient. The relationship between relative velocity and dry frictional coefficient is determined by the state of the contact surface between the bow and the string \cite{Bellante2015-si}. 

The physical properties of the strings are assumed to match those of the steel G strings used in the experiment: the reference frequency is $196.9\ {\rm Hz}$, the diameter is $0.8\ {\rm mm}$, and the length under tension but without load is $325\ {\rm mm}$. 
Young modulus and density values for steel are $200\ {\rm GPa}$ and $7700\ {\rm kg/m^3}$, respectively~\cite{Jansson2002Acoustics}. 
The simulation time step is $40\ {\rm \mu s}$ (equivalent to a sampling frequency of $25\ {\rm kHz}$), and the total simulation time is $0.26\ {\rm s}$. 
The position of the bow node is assumed to be 10\% of the string length from the bridge.
The pressure applied to the bowstring is controlled by the depth of the bow pressing depth, $\Delta z$.

This model enables the investigation of how the vibration spectrum depends on bow pressure and bowing speed.
Fig.~\ref{Fig:analysis}(b) shows the numerical simulation results obtained at the evaluation node represented by a double circle in Fig.~\ref{Fig:analysis}(a).
To remove the DC component, a Gaussian-like high-pass filter with a cutoff frequency of $f_0/4$ is applied to each spectrum. 
As shown in the upper plot, frequencies lower than $f_0$ are observed when the bow is pressed deeply. 
The $f_0/2$ peak is particularly significant for slow bow speed. 
As shown in the lower plot, harmonics appear when the bow depression is shallow. 
If the velocity of the bow is increased, the non-peak components disappear.
These simulation results are consistent with the experimental conditions for generating subharmonic frequencies.
The low-frequency modes shown in Fig.~\ref{Fig:analysis}(b) are clearly caused by the stick-slip motion, as confirmed by the similar characteristics observed in the vibration modes of the string in real space in both the experimental results and numerical calculations.

Obviously, the modes below the fundamental frequency shown in Fig.~\ref{Fig:analysis}(b) is caused by the slip-stick motion.
The vibration modes of the string in real space also show similar characteristics in the results of the experiment and numerical calculation.
Fig.~\ref{Fig:analysis}(c) shows the amplitude variation of the bow node under two conditions: when the bow is pushed in deeply and the speed is slow, and when it is pushed in shallowly and the speed is fast, as in (b). As in Fig.~\ref{Fig:exp}(c), we define $\tau\!=\!0$ as the moment when a typical slip occurrs.
The gray line represents the raw data, which contains fine structures attributed to energy accumulation due to the absence of a damping term in Eq.~\ref{eq:eom_no_bow2}. To suppress these fine structures, a low-pass filter with a half-width half-maximum window of 1 kHz is applied, and the filtered waveforms are plotted as red and blue solid lines.
In the subharmonic technique (blue solid line), the oscillation occurs around the position off from the original position compared to the normal case. 
The cycle of the basic slip-stick vibration is doubled in the subharmonic motion, excluding the subpeaks caused by the harmonic. 
Even with a simple model based on the tension between the elements and the friction function, our results demonstrate that subharmonic modes can be excited by adjusting the bow parameters. 

In conclusion, we experimentally visualized the two-dimensional string vibration modes during a performance of the subharmonic technique using high-speed imaging under actual performance conditions.
The captured spatial string deformations provided the first direct experimental evidence of Helmholtz motion reflecting at the bow-string contact point every two cycles during subharmonics, unambiguously demonstrating the physical mechanism responsible for generating frequencies below the fundamental frequency.

To further explore the physical origins of subharmonic generation, numerical FE simulations were conducted with frictional forces as the external driving mechanism.
The simulations reproduced the half-frequency behavior observed experimentally and confirmed that increased bow pressure and reduced bow speed play a major role in subharmonic production, reinforcing the experimental findings.

While further work is required, such as incorporating torsional and longitudinal waves into theoretical models~\cite{Guettler1994CASJ,Yoshikawa1997} and conducting parameter exploration experiments under controlled conditions using a monocode experimental setup, the combined use of experimental imaging and numerical modeling provides a robust framework for clarifying the mechanisms of complex vibrational phenomena.
Furthermore, the analogy between classical wave phenomena and electromagnetic waves has inspired research in both fields in recent years, with many parallels reported~\cite{Bliokh22,Peano15PRX}. A detailed understanding of empirically known special playing techniques in musical instruments may offer insights into nonlinear optical processes, such as frequency down-conversion.

\section{acknowledgments}
The authors thank M. Fujimura and M. Komuro for valuable discussions. The authors also thank K. Ichiji for critical contributions to high-speed imaging. The authors are grateful to R. Kita and T. Fukuda for their critical reading of the manuscript and insightful feedback. The authors sincerely appreciate N. Shitara for proofreading and providing valuable suggestions on the manuscript. S.K. thanks to the MERIT-WINGS program for a predoctoral fellowship. This work was supported by the Beseeltes Ensemble Tokyo.
\bibliography{sub}

\end{document}